%% file: main.tex
\documentclass[prl,twocolumn,nofootinbib,superscriptaddress,amsmath,amssymb]{revtex4-1}
\include{preamble}

\begin{document}

\title{Measuring Nuclear Matter Parameters with NICER and LIGO/Virgo}
\author{Josef Zimmerman}
\affiliation{%
 Department of Physics, University of Virginia, Charlottesville, Virginia 22904, USA
}%

\author{Zack Carson}
\affiliation{%
 Department of Physics, University of Virginia, Charlottesville, Virginia 22904, USA
}%

\author{Kristen Schumacher}
\affiliation{Department of Physics, University of Illinois at Urbana-Champaign, Urbana, IL, 61801, USA}

\author{Andrew W. Steiner}
\affiliation{%
 Department of Physics and Astronomy, University of Tennessee, Knoxville, TN 37996, USA
}%
\affiliation{%
 Physics Division, Oak Ridge National Laboratory, Oak Ridge, TN 37831, USA
}%

\author{Kent Yagi}
\affiliation{%
 Department of Physics, University of Virginia, Charlottesville, Virginia 22904, USA
}%

\date{February 8, 2020}

\pagecolor{white}

\begin{abstract}

The NICER Collaboration recently reported the measurement of the mass and radius of a pulsar PSR J0030+0451. We here use this new measurement to constrain one of the higher-order nuclear matter parameters $K_\sym$. 
We further combine the tidal measurement of the binary neutron star merger GW170817 by LIGO/Virgo to derive a joint 1-$\sigma$ constraint as $K_\sym = -102^{+71}_{-72}~\mev$. We believe this is the most reliable bound on the parameter to date under the assumption that there is no new physics above the saturation density which impacts neutron star observations.

\end{abstract}

\maketitle
\emph{Introduction}\label{sec:intro}---
The supranuclear equation of state (EoS), found in heavy ion collisions~\cite{Danielewicz:2002pu,Tsang:2008fd} and neutron stars (NSs)~\cite{Lattimer:2000nx}, remains one of the biggest mysteries in nuclear physics and astrophysics to date. 
The relationship between the NS's internal pressure and density is vital in determining macroscopic quantities such as the radius and tidal deformability.
As such, independent measurements of multiple NS observables can be used to constrain the EoS.
For example, x-ray measurements of the NS mass and radius have been used to probe the EoS~\cite{guver,ozel-baym-guver,steiner-lattimer-brown,Lattimer2014,Ozel:2016oaf}.

More recently, the Neutron Star Interior Composition Explorer (NICER) mounted on the International Space Station successfully mapped the NS mass-radius profile via x-ray observations of local hotspots rotating with the NS~\cite{Riley_2019,Miller_2019,Bogdanov_2019,Bogdanov_2019b,Guillot_2019,Raaijmakers_2019}.
Several follow-up investigations have further transformed such probability distributions of the NS radius with uncertainties on the order of $\sim10\%$ into constraints on the EoS~\cite{Raaijmakers_2019,Christian:2019qer,Jiang:2019rcw,Raaijmakers:2019dks}.

The GW detection of the coalescing binary NS system GW170817~\cite{TheLIGOScientific:2017qsa} by the LIGO/Virgo Collaborations (LVC) made history by opening a new window into the NS interior via tidal effects imprinted upon the gravitational waveform~\cite{Abbott2018,Abbott:2018exr,Paschalidis2018,Burgio2018,Malik2018,Landry:2018prl,Baiotti:2019sew,GuerraChaves:2019foa}.
Such tidal effects are a result of the tidal deformation each NS increasingly experiences in response to the neighboring inspiraling stars' tidal field and are characterized by the \emph{tidal deformability}~\cite{Flanagan2008}.
This instrumental measurement has been further mapped to constraints on the NS radius, the EoS and properties of quark matter~\cite{LIGO:posterior,Annala:2017llu,Abbott:2018exr,Lim:2018bkq,Bauswein:2017vtn,De:2018uhw,Most:2018hfd,Annala:2019puf}. 
GW and NICER measurements have been combined to constrain the EoS in~\cite{Raaijmakers:2019dks}.

In this letter, we combine the GW and x-ray observations from the LVC and NICER to strongly constrain the nuclear matter parameters indicative of the EoS.
The latter parameters are obtained by expanding binding energy per nucleon about nuclear saturation density and symmetric matter. 
Stemming from an early important work by Alam \emph{et al.}~\cite{Alam2016} where strong correlations were found between nuclear parameters and the NS radius, Refs.~\cite{Malik2018,Zack:nuclearConstraints,Carson:2019xxz} extended the analysis to find correlations between the former and the NS tidal deformability that have been applied to GW170817.
Here we construct, for the first time, 3-dimensional correlations between nuclear parameters, the NS radius, and the NS tidal deformability. We then use the GW and NICER measurements of the NS radius and tidal deformability to obtain joint GW+NICER constraints on the parameters, which are stronger than have been derived previously.

\par
\emph{Nuclear Matter parameters and Equations of State}\label{sec:eos}---
We begin by reviewing the nuclear matter parameters which we aim to measure through gravitational wave and x-ray observations in this letter. 
In order to quantify the properties of each EoS, we describe them using the parameters of a Taylor expansion in the space of nucleon number density $n$ and isospin symmetry parameter $\delta \equiv (n_n - n_p)/n$ in terms of the neutron ($n_n$) and proton ($n_p$) number densities. The parameter $\delta$ characterizes the amount of asymmetry in nuclear matter between neutrons and protons.

The expansion goes as follows.
We first express the energy per nucleon $e(n,\delta)$ as the sum of the symmetric matter part $e(n,0)$ plus the leading asymmetric part $S_2(n)$ as 
\begin{equation}
    e(n,\delta) = e(n,0) + S_2(n) \delta^2 + \mathcal{O}(\delta^4).    
    \label{eq:taylor1}
\end{equation}
We can further expand the symmetric part about nuclear saturation density $n_0$ using the parameters as
\begin{equation}
    e(n,0) = e_0 + \frac{K_0}{2} y^2 + \frac{Q_0}{6} y^3 + \mathcal{O}(y^4),
    \label{eq:taylor2}
\end{equation}
where $y \equiv (n-n_0)/3n_0$ and the coefficients represent energy per nucleon $e_0$, incompressibility $K_0$, and third derivative term $Q_0$, respectively. 
Similarly, we can expand the asymmetric part as
\begin{equation}
    S_2(n) = J_0 + L_0 y + \frac{K_\sym}{2} y^2 + \mathcal{O}(y^3),
    \label{eq:taylor3}
\end{equation}
where the coefficients represent symmetry energy $J_0$, its slope $L_0$, and its curvature $K_\sym$.
The lower order parameters in the expansion, such as $J_0$ and $L_0$, have been constrained with nuclear experiments~\cite{Tews2017}. On the other hand, neutron star observations can be used to measure higher order parameters like $K_\sym$ due to their large central densities. Indeed, Refs.~\cite{Malik2018,Zack:nuclearConstraints} used GW170817 to constrain this parameter.
In this letter, we exclusively focus on placing bounds on $K_\sym$ by combining results from GW170817 and the recent NICER measurements of the neutron star radius.

We consider a population of EoSs of various types to encompass the correlations with observable NS properties: 14 Skyrme-type, 5 relativistic mean field (RMF), and 38 phenomenological EoSs (PE). 
The first two classes were studied in~\cite{Alam2016,Malik2018}, while the latter class was generated in~\cite{Carson:2019xxz} by uniformly sampling the six nuclear parameters and modeling it exactly as the form in Eqs.~\eqref{eq:taylor1}-\eqref{eq:taylor3}.
We excluded EoSs that allowed acausal speed of sound (the speed of sound exceeding the speed of light) or decreasing pressure with increasing density.
Additionally, we rejected EoSs inconsistent with the 90\% confidence bounds on $L_0$ and $J_0$ described in~\cite{Tews2017}. 
All of the selected EoSs support NSs with mass above $\sim 2 M_\odot$~\cite{1.97NS,2.01NS,Cromartie:2019kug}.
The resulting population of EoSs encompasses the physically motivated Skyrme and RMF models, and the inclusion of phenomenological models increases the breadth of our models, minimizing systematic errors from choice of models, distinguishing our work from previous studies such as that of~\cite{Alam2016}, which did not consider PE models. All the EoS models considered here are nucleonic ones and do not contain hyperons, Bose condensates nor quarks.

\begin{figure}
    \centering
    \includegraphics[width=\columnwidth]{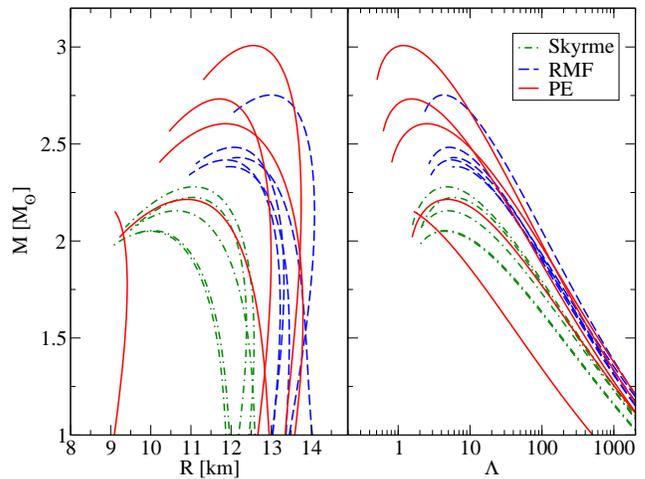}
    \caption{Relations between the NS mass and radius (left) or tidal deformability (right) for three different classes of EoSs. 
    }
    \label{fig:EoS}
\end{figure}

\emph{Mass, Radius and Tidal Deformability}---
We use the above EoSs to compute NS observables, such as mass, radius and tidal deformability. One can extract the first two quantities by constructing a non-rotating, isolated NS solution by solving the Einstein equations (or Tolman-Oppenheimer-Volkhoff equations). We solve these differential equations by assuming a central density or pressure as an initial condition. One then extracts the mass from the asymptotic behavior of the gravitational potential, while the radius is determined as the location of vanishing pressure. The left panel of Fig.~\ref{fig:EoS} presents the mass-radius relation for selected EoSs. Observe that Skyrme  EoSs produce NSs with smaller radii, while RMF ones produce NSs with larger radii, and the phenomenological EoSs cover a wide range of the parameter space.

The tidal deformability $\lambda$ of a NS quantifies its elasticity to develop a quadrupole moment $Q_{ij}$ in the presence of an external tidal field $\mathcal{E}_{ij}$ as 
\begin{equation}
    Q_{ij} = -\lambda \ \mathcal{E}_{ij}.
\end{equation}
$Q_{ij}$ and $\mathcal{E}_{ij}$ are both obtained from the asymptotic behavior of the gravitational potential around a tidally-deformed NS. Such a stellar solution is constructed by perturbing the non-rotating, isolated background solution derived earlier and solving a set of perturbed Einstein equations. The right panel of Fig.~\ref{fig:EoS} presents the relation between the mass and the dimensionless tidal deformability $\Lambda \equiv \lambda / M^5$. Skyrme and RMF EoSs produce NSs with smaller or larger tidal deformabilities, respectively, while phenomenological EoSs cover a wide region of the parameter space, similar to the mass-radius relation in the left panel. 

As tidal interactions arise from two individual NSs, it is difficult to independently measure the tidal deformability of each NS by itself due to strong correlations between the two.
Rather, we consider the dominant tidal parameter in the gravitational waveform from binary neutron star mergers, which corresponds to the mass-averaged tidal deformability:
\begin{equation}
	\tilde{\Lambda} = \frac{16}{13} \frac{(1+12 q)\Lambda_1 + (12 + q)q^4 \Lambda_2}{(1+q)^5}.
    \label{eq:lambdaTilde}
\end{equation}
Here $q \equiv m_2/m_1 < 1$ is the mass ratio with $m_A$ and $\Lambda_A$ representing the mass and dimensionless tidal deformability of the $A$th neutron star respectively.

In~\cite{Zack:nuclearConstraints}, we derived constraints on $K_\sym$ as follows (see also~\cite{Malik2018}). We first studied the correlation between $K_\sym$ and $\tilde \Lambda$ using the three classes of the EoSs described earlier. Based on this, we constructed a 2D Gaussian probability distribution on these two parameters. We then multiplied this with the posterior probability distribution of $\tilde \Lambda$ obtained from GW170817 by the LIGO/Virgo Collaborations~\cite{LIGO:posterior}. We finally marginalize over $\tilde \Lambda$ to find the probability distribution of $K_\sym$.  Below, we follow a similar procedure to derive bounds on $K_\sym$ from NICER by using the correlation between such nuclear parameter and the NS radius that was first studied in~\cite{Alam2016}. 

\begin{figure}
    \centering
    \includegraphics[width=\columnwidth]{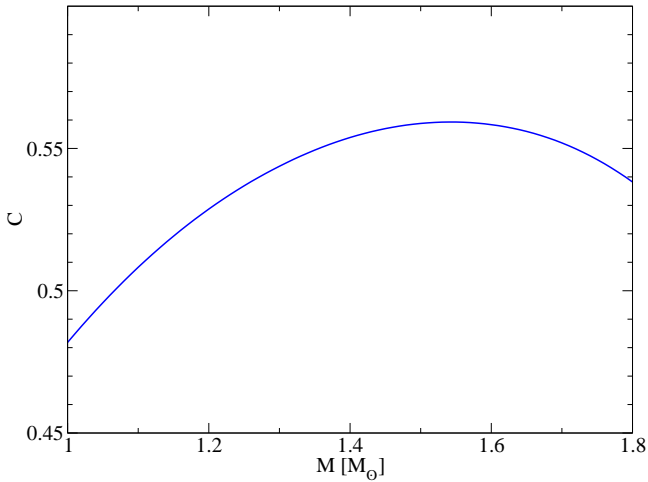}
    \caption{Correlation between $K_\sym$ and $R$ as a function of the NS mass.}
    
    \label{fig:correlation}
\end{figure}

%correlations and gaussian
\emph{Correlation between Nuclear Matter and Neutron Star Parameters}\label{sec:gauss}---
To map measurements of neutron star quantities to those for nuclear matter parameters, we need to study correlations between them. For example, the correlation between $K_\sym$ and $\tilde \Lambda$ were studied in~\cite{Zack:nuclearConstraints,Carson:2019xxz}. In a similar manner, we present the correlation between $K_\sym$ and $R_{1.4}$ (the NS radius with mass $1.4 M_\odot$) for the three classes of EoSs. Observe that there is a strong correlation between the two quantities, which can be quantified by the Pearson correlation coefficient~\cite{Alam2016}:
\begin{equation}
	C = \frac{\Sigma_{xy}}{\sqrt{\Sigma_{xx}\Sigma_{yy}}}.
	\label{equation:correlation}
\end{equation}
Here $\Sigma_{xy}$ are covariance matrix elements between parameters $x$ and $y$ given by
\begin{equation}
	\Sigma_{xy} = \frac{1}{N}\sum_{i=0}^{N} x_i y_i - \frac{1}{N^2}\left(\sum_{i=0}^{N} x_i \right)\left(\sum_{i=0}^{N} y_i\right),
	\label{equation:covariance}
\end{equation}
with $N$ representing the number of data points. Figure~\ref{fig:correlation} presents the correlation coefficient for various NS masses. Notice that the correlations are around 0.5, which is weaker than that found in~\cite{Alam2016} using only the Skyrme and RMF classes of EoSs. Thus, our results that include phenomenological EoSs are more conservative as they account for more variation and capture the trends of correlation more accurately. 

\begin{figure}
    \centering
    \includegraphics[width=\columnwidth]{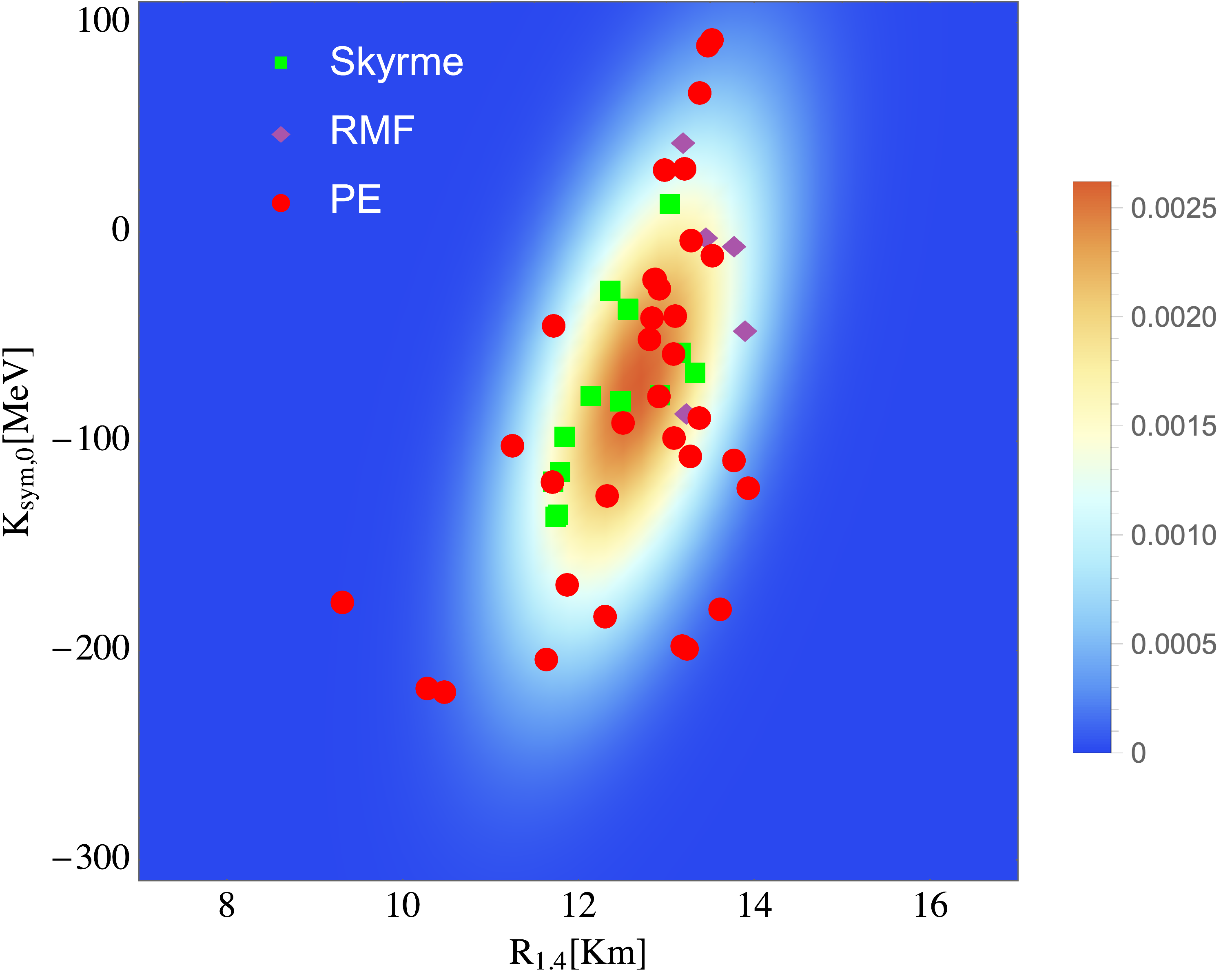}
    \caption{Correlation between $K_\sym$ and the NS radius with mass 1.4$M_\odot$. We show the scattering with Skyrme (green), RMF (purple) and PE (red) EoSs. The colored contour shows the two-dimensional Gaussian distribution fit.
    }
    \label{fig:ksym_correlation}
\end{figure}

Based on this correlation, we construct a two-dimensional Gaussian distribution given by
\begin{equation}
    P(\bm{x}) = \frac{1}{\sqrt{(2 \pi)^N |\bm{\Sigma}|}} e^{-\frac{1}{2}(\bm{x}-\bm{\mu})^T\bm{\Sigma}^{-1}(\bm{x}-\bm{\mu})}, 
    \label{equation:multiGauss}
\end{equation}
with $N=2$ representing the number of variables  and $\bm \mu$ corresponding to the mean of each variable. $\bm \Sigma$ is the covariance matrix already defined in~\eqref{equation:covariance}. Figure \ref{fig:ksym_correlation} also shows this Gaussian fit in colored contours. We repeat the analysis for various NS masses $M$ and create yet another fit for $\bm \mu$ and $\bm \Sigma$ of the form
\begin{equation}
	a e^{b m + c m^2 + d m^3}
	\label{equation:fitForm}
\end{equation}
with $m \equiv M/M_\odot$.
The fitted coefficients $a$, $b$, $c$ and $d$ are summarized in Table~\ref{table:gaussFitsSigmas}.

We next derived a three-dimensional correlation among $K_\sym$, $R_{1.4}$ and $\tilde \Lambda_{1.186}$, where $\tilde \Lambda_{1.186}$ refers to $\tilde \Lambda$ with a chirp mass of ${\mathcal M} = 1.186~M_\odot$ corresponding to GW170817~\cite{Abbott:LTposterior}\footnote{The chirp mass is defined as $\mathcal M \equiv (m_1^3 m_2^3/M_\mathrm{tot})^{1/5}$ with the total mass $M_\mathrm{tot} = m_1+m_2$. The mass ratio has a larger uncertainty than the chirp mass, though the correlation between $K_\sym$ and $\tilde \Lambda$ is insensitive to the mass ratio~\cite{Zack:nuclearConstraints}.}. We then created a three-dimensional Gaussian fit following Eq.~\eqref{equation:multiGauss} with $N=3$, and then created another fit for $\bm \mu$ and $\bm \Sigma$ following Eq.~\eqref{equation:fitForm}. The fitted coefficients are again listed in Table~\ref{table:gaussFitsSigmas}.

\renewcommand{\arraystretch}{1.2}
\begin{table*}[ht]
    \centering
        \begin{tabular}{c|c c|c c c c}
              &  $x$ & $y$ & $a$ & $b$ & $c$ & $d$ \\
                \hline \hline
                 &$R_{\M}$ & --- &13.03 km & -0.1153 & 0.1145 & -0.0333 \\
     $\mu_x$   \    & $K_\sym$  &--- & -72.91 MeV & --- & --- & ---\\
              &   $\tilde{\Lambda}_{1.186}$ &--- & 699.2 & --- & --- & --- \\
                \hline
       \multirow{6}{*}{$\Sigma_{xy}$}   \    &   $R_{\M}$ & $R_{\M}$  & 1.756 km$^2$ & -0.4836 & -0.3745 & 0.2420\\
              &   $K_\sym$ & $K_\sym$  & 6382 MeV$^2$& --- & --- & --- \\
              &   $K_\sym$ &$R_{\M}$  & 16.66 /km$\cdot$ MeV & 1.448 & -0.7716 & 0.1350\\
              &   $R_{\M}$ & $\tilde{\Lambda}_{1.186}$ &  217.4 km & $\sci{-4.922}{-2}$ & $\sci{-5.417}{-2}$ & $\sci{4.437}{-2}$\\
              &   $K_\sym$ &$\tilde{\Lambda}_{1.186}$ & $\sci{1.270}{4}$ MeV & --- & --- & --- \\
              &   $\tilde{\Lambda}_{1.186}$ & $\tilde{\Lambda}_{1.186}$ & $\sci{5.654}{4}$ & --- & --- & ---\\
        \end{tabular}
    %}
        \caption{Fitted coefficients in Eq.~\eqref{equation:fitForm} for the mean $\bm \mu$ and covariance $\bm \Sigma$ of the two-dimensional (the ones that do not involve $\tilde \Lambda_{1.186}$) and three-dimensional Gaussian distributions given by Eq.~\eqref{equation:multiGauss}.}
    \label{table:gaussFitsSigmas}
\end{table*}

\emph{Bounds on $K_\sym$ from NICER}---
We now derive bounds on $K_\sym$ from the NICER's measurement of the NS mass and radius for PSR J0030+0451.
We explore the results of the NICER study~\cite{Miller_2019}, released as the Monte Carlo Markov Chain (MCMC) samples for pairs of mass and radius measured from the NS.
We first divide NICER's results into 20 bins of mass evenly spaced between 1 and 1.95 $\msol$\footnote{We have checked that when we increase the number of bins to 40, the final bound on $K_\sym$ only changed by less than 1 MeV.}.  The MCMC samples within each mass bin that starts with mass $M$ give us the probability distribution of the radius at $M$: $P(R_\M)$.
Below, we mainly use the NICER results for the 3-spot model, though we found that using the 2-spot model does not affect our final bound much.

We next use marginalization integrals to combine measurements of $R$ and correlations in the Gaussian distributions to extract bounds on $K_\sym$. 
The probability distribution of $K_\sym$ from one mass bin at $M$ is given by
\begin{equation}
    P_\M(K_\sym) = \int_{-\infty}^{\infty} P_\M(K_\sym,R_\M) \, P(R_\M) \, dR_\M.
    \label{eq:mar1}
\end{equation}
where $P_\M(K_\sym,R_\M)$ is the two-dimensional Gaussian distribution between $K_\sym$ and $R_\M$ constructed by Eq.~\eqref{equation:multiGauss}. This way, we can take into account the amount of scattering in the correlation between $K_\sym$ and $R_\M$, which adds a systematic error to the final distribution on $K_\sym$.
After normalizing it properly, this yields bounds on $K_\sym$ for each fixed value of mass $M$.
The final step combines each bound weighted by the probability distribution of the mass $P(M)$ constructed from the number of NICER samples in each mass bin: 
\begin{equation}
    P(K_\sym) = \sum P_\M(K_\sym) P(M).
    \label{eq:mar2}
\end{equation}

\begin{figure}
    \centering
    \includegraphics[width=\columnwidth]{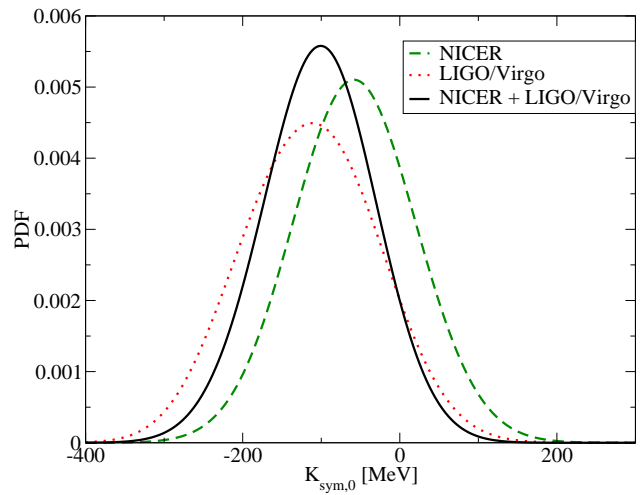}
    \caption{Comparison of probability distributions for $K_\sym$ with various observations: PSR J0030+0451 with NICER using the 3-spot model, 
    GW170817 with LIGO/Virgo,
    and NICER + LIGO/Virgo combined. }
    \label{fig:ksym_all}
\end{figure}

The resulting probability distribution of $K_\sym$ is shown in Fig.~\ref{fig:ksym_all}. For comparison, we also present the distribution using GW170817 with LIGO/Virgo~\cite{Zack:nuclearConstraints}. Observe that the NICER result gives us a slightly tighter bound than that from LIGO/Virgo.

To place more stringent bounds on $K_\sym$, we now combine the two measurements from NICER and LIGO/Virgo.
%NICER results with previous NS measurements from GW170817.
We use the three-dimensional probability distribution $P_\M(K_\sym, R_\M, \tilde{\Lambda}_{1.186})$ from Eq.~\eqref{equation:multiGauss} relating $K_\sym$, $R_\M$ and $\tilde \Lambda_{1.186}$. We can first use the tidal measurement of $P(\tilde{\Lambda}_{1.186})$ by LIGO/Virgo~\cite{LIGO:posterior} to marginalize the above three-dimensional probability distribution over $\tilde \Lambda_{1.186}$ to obtain $P_\M(K_\sym,R_\M)$:
\begin{eqnarray}
    %P_\M(K_\sym,R_\M,\tilde{\Lambda}_{1.186}) &=& \int_{-\infty}^\infty P_\M(K_\sym,R_\M,\tilde{\Lambda}_{1.186}) \nonumber \\
    P_\M(K_\sym,R_\M) &=& \int_{-\infty}^\infty P_\M(K_\sym,R_\M,\tilde{\Lambda}_{1.186}) \nonumber \\
    && \times P(\tilde{\Lambda}_{1.186})\, d \tilde{\Lambda}_{1.186}.
    \label{eq:mar3}
\end{eqnarray}
We then proceed according to Eqs.~\eqref{eq:mar1} and~\eqref{eq:mar2} to obtain final bounds on $K_\sym$, which we present in Fig. \ref{fig:ksym_all}. Observe that the bounds from the joint analysis of NICER and LIGO/Virgo is stronger than those from NICER or LIGO/Virgo alone. We summarize all the bounds in Table~\ref{table:summary} (including those from the 2-spot model). 

Other works have also obtained tight constraints on 
$K_\sym$, but employ additional 
assumptions. Reference~\cite{Dong2012} obtains a range of $K_\sym = -125\pm79~\mev$, by fixing
$J_0=31.6\pm2.2~\mathrm{MeV}$, but this upper limit is not consistent with recent theory
results~\cite{Kruger13}. Reference~\cite{Xie:2019sqb} finds a range $K_\sym = -230^{+90}_{-50}~\mev$,
but this range is based, in part, on a constraint from quiescent low-mass x-ray binaries (QLMXBs) of $R_{1.4}=11.7\pm 1.1~\mathrm{km}$
from Ref.~\cite{Lattimer2014}. A more recent work with a more complete 
analysis of QLMXB systematic uncertainties~\cite{Steiner:2017vmg} gives a much weaker 
constraint on the neutron star radius.
On the other hand, the NS radius measurement with NICER is expected to suffer from less systematics~\cite{Lo:2013ava}.

\renewcommand{\arraystretch}{1.4}
\begin{table}[ht]
    \centering
        \begin{tabular}{c|c c}
              &  NICER & \quad NICER + LIGO/Virgo \\
                \hline 
           2-spot   & $-47^{+81}_{-80}$ & $-102^{+71}_{-73}$\\
           3-spot   & $-58^{+79}_{-78}$ & $-102^{+71}_{-72}$\\
        \end{tabular}
    %}
        \caption{1-$\sigma$ Bounds on $K_\sym$ in units of MeV to $1\sigma$ confidence from NICER and NICER + LIGO/Virgo. We show results from both 2-spot and 3-spot models with NICER. For reference, the bound from LIGO/Virgo alone is $-114^{+87}_{-88}~\mev$~\cite{Zack:nuclearConstraints}.
        }
        
    \label{table:summary}
\end{table}

\emph{Conclusions and Discussions}---
We used the correlations between $K_\sym$ and $R_\M$ and have successfully mapped NICER's radius measurement to bounds on $K_\sym$. Moreover, we used a three-dimensional correlation among the above two parameters plus $\tilde \Lambda$ to find another probability distribution on $K_\sym$ by combining NICER's radius measurement with the tidal measurement of LIGO/Virgo. We derived the bound on $K_\sym$ that, we believe, is more reliable than previous bounds~\cite{Dong2012,Xie:2019sqb}, under the assumption that quarks and other phase transitions are absent beyond the saturation density.

Future work includes considering NS radius measurements from other X-ray observations, e.g. for quiescent low-mass X-ray binaries~\cite{Steiner:2017vmg}, to further strengthen bounds on $K_\sym$. Such radius measurements may suffer from large systematic errors, which need to be taken into account when deriving bounds on $K_\sym$. Work along this direction is currently in progress~\cite{Zimmerman_prep}. 

\emph{Acknowledgments}\label{sec:acknowledgements}---
A.W.S. was supported by NSF grant PHY 19-09490 and by the U.S. DOE Office of Nuclear Physics.
Z.C. and K.Y. acknowledge support from the Ed Owens Fund.
K.Y. also acknowledges support from NSF Award PHY-1806776, a Sloan Foundation Research Fellowship, the COST Action GWverse CA16104 and JSPS KAKENHI Grants No. JP17H06358.

\bibliography{main}

\end{document}

%% file: preamble.tex
\usepackage{mathtools}
\usepackage{physics}
\usepackage{graphics}
\usepackage{graphicx}
\usepackage{dcolumn}
\usepackage{bm}
\usepackage{dsfont} 
\usepackage{amsmath,amssymb}
\usepackage{hyperref}
\usepackage{tabularx}
\usepackage{epsf,epsfig}
\usepackage[normalem]{ulem}
\usepackage[usenames]{color}
\usepackage{multirow}
\usepackage{makecell}
\usepackage{diagbox}
\usepackage[abs]{overpic}
\allowdisplaybreaks
\hypersetup{
    colorlinks=true,
    linkcolor=blue,
    filecolor=magenta,      
    urlcolor=blue,
    citecolor=blue
}
\usepackage{pagecolor}
\usepackage{bm}

\newcommand{\M}{{\mbox{\tiny M}}}

\definecolor{red(ncs)}{rgb}{0.77, 0.01, 0.2}
\definecolor{utk}{rgb}{1.0, 0.51, 0.0}

\newcommand{\sci}[2]{{#1 \times 10^{#2}}}
\newcommand{\msol}{M_{\odot}}
\newcommand{\mev}{\mathrm{MeV}}
\newcommand{\sym}{\mathrm{sym,0}}

\usepackage{array}
\newcolumntype{C}[1]{>{\centering\let\newline\\\arraybackslash\hspace{0pt}}m{#1}}

\newcolumntype{C}[1]{>{\centering\arraybackslash}m{#1}}